\documentclass[prd,reprint,nofootinbib,showpacs,superscriptaddress]{revtex4-1}
\usepackage{graphicx} 
\usepackage{hyperref}
\usepackage{amsfonts}
\usepackage{amsmath,amssymb}
\usepackage{bm} 
\usepackage{color}
\usepackage{epstopdf} 
\usepackage{epsfig}
\usepackage{subfig} 

{\rm }

\def\be{\begin{equation}}
 \def\ee{\end{equation}}
 \def\bea{\begin{eqnarray}}
 \def\eea{\end{eqnarray}}
 \def\bes{\begin{eqnarray}}
 \def\ees{\end{eqnarray}}
 \def\bi{\begin{itemize}}
 \def\ei{\end{itemize}} 

\def\2{\frac{1}{2}}
\def\4{\frac{1}{4}}

\begin{document}

\title{Trustworthiness of detectors in quantum key distribution with untrusted detectors
}

\author{Bing Qi}
\email{qib1@ornl.gov}
\affiliation{Quantum Information Science Group, Computational Sciences and Engineering Division,
Oak Ridge National Laboratory, Oak Ridge, TN 37831-6418, USA}
\affiliation{
 Department of Physics and Astronomy, The
University of Tennessee, Knoxville, TN 37996 - 1200, USA
}

\date{\today}

\begin{abstract}

Measurement-device-independent quantum key distribution (MDI-QKD) protocol has been demonstrated as a viable solution to detector side-channel attacks. One of the main advantages of MDI-QKD is that the security can be proved without making any assumptions about how the measurement device works.  The price to pay is the relatively low secure key rate comparing with conventional QKD, such as the decoy-state BB84 protocol.  To bridge the strong security of MDI-QKD with the high efficiency of conventional QKD, a new type of QKD protocol, the detector-device-independent (DDI) QKD has been proposed recently. In this protocol, the legitimate receiver employs a trusted linear optics network to encode information on photons received from an insecure quantum channel, and then performs a single-photon Bell state measurement (BSM) using untrusted detectors.  One crucial assumption made in DDI-QKD is that the untrusted BSM located inside the receiver's laboratory cannot send any unwanted information to the outside. Here, we show that if the BSM is completely untrusted and the expected detection efficiency is low, a simple scheme would allow the BSM to send information to the outside: a sophisticated eavesdropper (Eve) can place high-efficiency detectors inside the BSM and program it to selectively report a fraction of the total detection events to simulate low-efficiency detectors. The time delay between adjacent reported events can be used by the BSM to send information. Combined with Trojan horse attacks, this scheme could allow Eve to gain information of the quantum key without being detected. To prevent the above attack, either countermeasures to Trojan horse attacks or some trustworthiness to the ``untrusted'' BSM device is required.   

\end{abstract}

\maketitle

\section{Introduction}

Quantum key distribution (QKD) allows two authenticated users, normally referred to as Alice and Bob, to generate a private key through an insecure quantum channel controlled by an eavesdropper, Eve \cite{BB84,E91,Gisin02,Scarani09,Lo14}. Based on fundamental laws in quantum mechanics, idealized QKD protocols have been proved to be unconditionally secure against adversaries with unlimited computing power and technological capabilities \cite{Mayers01,Lo99,Shor00}. However, practical implementations of QKD protocols unavoidably contain imperfections which may be overlooked in the security proofs.  The disconnection between QKD theory and its real-life implementations has led to various ``side-channel'' attacks \cite{Makarov05,Makarov06,Qi07,Zhao08,Lydersen10,Wiechers11,Lydersen102,Gerhardt11,Fung07,Xu10,Sun11,Tang13,Rogers07,Viacheslav14,Weier11,Li11}.

One natural solution to the above side-channel attacks is to develop security patches \cite{Yuan11,Lydersen11,Yuan112,Honjo13,Silva14}. In fact, once a security loophole has been identified, normally it is not too difficult to develop countermeasures to regain the security. This approach is appealing in practice since it typically can be realized by slightly modifying either the QKD protocol or its implementation, and the resulting key rate could be comparable with that of the original protocol.  However the security can only be proved against known attacks.

Another important approach to enhance the security of practical QKD is to develop QKD protocols based on ``untrusted'' device \cite{Mayers98,Acin07,Gisin10,Lo12,Braunstein12}. Among them, the measurement-device-independent (MDI) QKD protocol \cite{Lo12}, has received much attention \cite{Tamaki12,Ma12,Song12,Sun13,Xu13,Curty14}. The MDI-QKD protocol is automatically immune to all side-channel attacks associated with the measurement device which, arguably, is the weakest link in a QKD system. In fact, the measurement device in MDI-QKD can be treated as a ``black box'' which could even be manufactured and operated by Eve. The feasibility of MDI-QKD has been demonstrated experimentally \cite{Rubenok13,Silva13,Liu13,Tang14,Tang142,Tang143}. See \cite{Xu14} for a recent review.

One drawback of MDI-QKD is the relatively low secure key rate comparing with conventional QKD, such as the decoy-state BB84 protocol. This is because in most of existing MDI-QKD protocols (except \cite{Tamaki12}), secure keys are generated from two-fold coincidence detection events. The MDI-QKD suffers more from the low detection efficiency of a practical single-photon detector (SPD), especially when the finite data size effect is taken into account \cite{Lo14}.
 
Recently a new QKD protocol, designed to bridge the strong security of MDI-QKD with the high efficiency of conventional QKD, was proposed by several groups \cite{Gonzalez14,Lim14,Cao14}. In this protocol, the legitimate receiver employs a trusted linear optics network to encode information on photons received from an insecure quantum channel, and then performs a Bell state measurement (BSM) using untrusted detectors. One important advantage of this approach is the high key rate comparable with that of a conventional QKD protocol. This is achieved by replacing the probabilistic two-photon BSM scheme in the original MDI-QKD protocol \cite{Lo12} by a deterministic single-photon BSM scheme. Following \cite{Lim14}, we call this new protocol ``detector-device-independent'' (DDI) QKD.

In this paper, we scrutinize the underlying assumptions behind DDI-QKD. One crucial assumption is that the untrusted BSM located inside the receiver's laboratory cannot send any ``unwanted'' information to the outside. Here, we show that if the BSM is completely untrusted, a simple scheme would allow the BSM to send information to the outside without being detected: Eve can place high-efficiency detectors inside the BSM and program it to selectively report a fraction of the total detection events to simulate practical low-efficiency detectors. The time delay between adjacent reported events can be used by the BSM to send information. Combined with Trojan horse attacks, this scheme could allow Eve to gain information of the secure key without introducing any errors. Our results suggest that to establish the security of DDI-QKD, additional assumptions on the measurement device are required. It is thus very important to clearly spell out those assumptions and place them under scrutiny. 

This paper is organized as follows: in Section \ref{sec:2}, we briefly review the basic ideas of MDI-QKD and DDI-QKD.  In Section \ref{sec:3}, we present the details of our attack. We conclude this paper with a discussion in Section \ref{sec:4}.

\section{MDI-QKD \& DDI-QKD}
\label{sec:2}

In a conventional QKD protocol (see Fig.1a), Alice prepares quantum states and sends them to Bob through an insecure quantum channel. Bob receives signals from the quantum channel and performs measurement. In this configuration, it is reasonable to assume that the errors in quantum state preparation can be well controlled and quantified, since this can be done within Alice's well protected laboratory without Eve's interference. On the other hand, the quantum states received by Bob are highly unpredictable, since Eve could replace Alice's quantum states by anything at her will. Eve can interfere the measurement process by either manipulating Alice's signal \cite{Qi07} or sending her own signals into Bob's laboratory \cite{Lydersen102}. The above observation could explain why most identified security loopholes in conventional QKD are associated with the measurement device \cite{Makarov05,Makarov06,Qi07,Zhao08,Lydersen10,Wiechers11,Lydersen102,Gerhardt11}.

\begin{figure}[htb]
\centerline{\includegraphics[width=8.5cm]{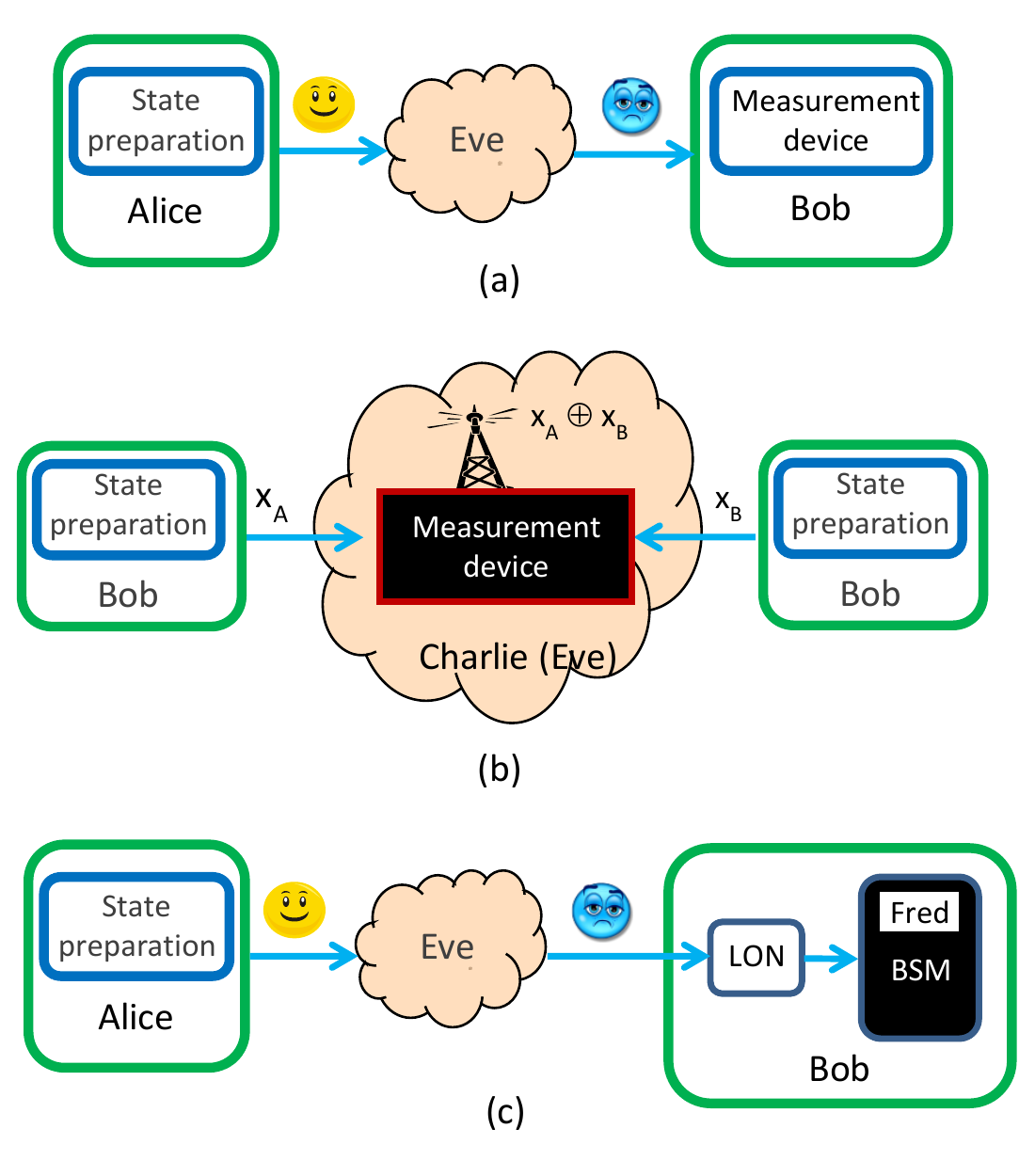}} \caption{(a) Conventional QKD; (b) Measurement-device-independent(MDI) QKD; (c) Detector-device-independent (DDI)QKD}
\end{figure}

The motivation behind MDI-QKD \cite{Lo12} is to develop a QKD protocol which is automatically immune to all detector side-channel attacks. In this scheme (see Fig.1b), both Alice and Bob prepare quantum states and send them to an untrusted third party, Charlie, who could be a collaborator of Eve. Charlie is supposed to measure the correlation between Alice's and Bob's quantum states and publicly announce the measurement results. Given Charlie's measurement results, Alice and Bob can further establish a secure key. The protocol has been designed in such a way that ``only'' the correlation (but not the quantum states themselves) can be determined by Charlie faithfully. On one hand, if Charlie executes the protocol honestly, he or Eve cannot gain any information of the secure key. On the other hand, any attempts by Charlie to gain information of the secure key will unavoidably introduce additional noise and can be detected by Alice and Bob. The security of MDI-QKD is based on the idea of time-reversed EPR QKD \cite{Biham96,Inamori02}. By allowing Eve to fully control the measurement device, Alice and Bob can establish a secure key without making any assumptions about how the measurement device works, thus removing any potential detector side-channels. We remark that MDI-QKD is an ideal building block of multi-user QKD networks, where the expensive measurement device can be placed in an untrusted central node while each QKD user only needs a low-cost state preparation device \cite{Tang14}. 

Recently DDI-QKD has been proposed to bridge the strong security of MDI-QKD with the high efficiency of conventional QKD \cite{Gonzalez14,Lim14,Cao14}. For simplicity, in this paper we assume that a perfect single photon source is employed by Alice. In DDI-QKD (Fig.1c), Alice prepares a single-photon pulse in one of the four BB84 polarization states and sends it to Bob through an insecure quantum channel. On receiving the signals, Bob employs a ``trusted'' linear optics network (LON) to encode his information on a different (for example, spatial) degree of freedom of the incoming signals. Afterwards, the above signals are fed into an ``untrusted'' BSM device which is supposed to perform a single-photon BSM \cite{Kim03} and report the measurement results to Bob. Through an authenticated classical channel, Bob announces which transmitted signals have yielded successful BSM detection events, the Bell states obtained, and the basis information associated with those successful events.  Alice and Bob estimate the quantum bit error rate (QBER) for the events when they happen to use the same basis. If the QBER is below certain threshold, they can further apply error correction and privacy amplification to generate a secure key. Note in Fig.1c, the untrusted BSM inside Bob's laboratory is represented by a ``black box'', which is controlled by Eve's partner (Fred) during the QKD process.

\section{Trojan horse attack in DDI-QKD}
\label{sec:3}

In DDI-QKD, Bob's encoding device (LON in Fig.1c) sits between Eve and Fred. In practice, this design could be prone to Trojan horse attacks. We remark that the threats of Trojan horse attacks in conventional QKD and the corresponding countermeasures have been studied previously \cite{Gisin06,Jain14,Jain142}. In conventional QKD, Eve can only access one end of each user's device. To launch a Trojan horse attack, Eve can send bright light pulses into the user's system and tries to gain information by measuring the back-reflected light. In practice, the QKD users can effectively reduce the risk of Trojan horse attack by using filters, optical circulators, isolators, and intensity monitors, etc. However, these countermeasures may not be effective in the case of DDI-QKD, where Eve and Fred together can access both ends of Bob's encoding device. For example, in each quantum transmission, Eve could send her own signal (which may contain a few photons) together with Alice's photon into Bob's laboratory. Both Alice's and Eve's signals go through Bob's encoding device and reach the BSM. Inside the BSM, Fred could determine Bob's bit information precisely by measuring the signal sent by Eve. In the meantime Fred can perform an honest Bell state measurement on Alice's photon. In principle, Fred can have a perfect copy of Bob's random bits without introducing any errors.

At this point, security is not compromised since Fred is confined within Bob's laboratory.  To prevent Fred from sending Bob's random bits to Eve, a crucial assumption is made in DDI-QKD \cite{Gonzalez14,Lim14}: Fred is only allowed to report the BSM results to Bob; he cannot send any ``unwanted'' information to the outside. However, it could be difficult to justify the above assumption in practice. Below we will show that if the BSM is completely untrusted (as in the case of MDI-QKD), a simple scheme would allow the untrusted BSM to send information to Eve. Combined with Trojan horse attacks discussed above, this scheme could allow Eve to gain information of the final key without being detected.

A practical single photon detector (SPD) at telecom wavelength has a relatively low detection efficiency (typically $10\%\sim30\%$). On the top of that, all the optical components inside a practical BSM introduce additional losses. Under normal operation, Bob would expect a low detection rate: most of the time, the BSM will report ``no detection''; occasionally, the BSM will report a successful BSM result. 
However, Eve could place SPDs with much higher efficiency inside the BSM. In this case, the actual detection rate seen by Fred will be much higher than the one expected by Bob. Fred can easily simulate a low-efficiency BSM by reporting to Bob a small fraction of the total detection events. He can further take advantage of this ``post-selection'' process to send information to Eve.

Suppose Fred successfully detects the $i^{th}$ signal sent by Alice; in the meantime, through the Trojan horse attack, he also learns Bob's bit information corresponding to the same transmission. Fred reports the BSM result to Bob honestly. He also determines the index number $i+k$ of the next BSM result to be reported based on the following rules: if Bob's $i^{th}$ bit value is $1$ (or $0$), Fred will report a BSM result to make sure that $k$ is an even (or odd) number. In other words, Fred encodes Bob's random bits on the time delays between adjacent BSM results reported to Bob. When Bob publicly announces which signals from Alice have been detected, Eve can decode Fred's information and have a perfect copy of Bob's random bits. Since Fred performs BSM on Alice's signals and reports the measurement results to Bob honestly, this attack will not introduce additional errors. To further conceal their attack, Eve and Fred can use pre-shared random numbers to determine whether an even or an odd number is used to encode bit $1$ for each reported BSM event. Fred can also carefully control the reporting rate to match it with the one expected by Bob.

This attack, in its spirit, is similar to the memory attack on device-independent (DI) QKD \cite{Barrett13}, which might be applicable whenever an untrusted device is placed inside Alice's or Bob's secure laboratory. On the other hand, our attack is more feasible since Fred does not need to access the classical communication channel between Alice and Bob. All his activities are confined within the untrusted BSM, as assumed in DDI-QKD.

\section{Discussion}
\label{sec:4}

To prevent the above attack, Bob could introduce various countermeasures to detect Trojan horse attacks, as suggested in \cite{Cao14}. In practice, additional filtering and random sampling systems could be introduced into Bob's system to characterize the input signals and mitigate the risk of Trojan horse attacks. However, it is very challenging in practice to implement single-mode filtering. Moreover, additional SPDs could be required to implement the above countermeasures. These SPDs may also suffer from the same side-channel attacks as the SPDs for secure key generation.

Another approach to prevent the above attack is to make additional assumptions about the BSM device \cite{Curty}. Instead of treating the BSM device as a completely ``black box'', as we have assumed above, Bob could build the BSM himself and know what is inside the BSM. In this case, it may be possible to prove security without perfectly modeling the exact behavior of the BSM. Nevertheless, all the assumptions made about the BSM should be clearly specified and be placed under scrutiny. To highlight this point, we consider the detector blinding attack, which has been successfully demonstrated in conventional QKD systems \cite{Lydersen102}.

\begin{figure}[htb]
\centerline{\includegraphics[width=7cm]{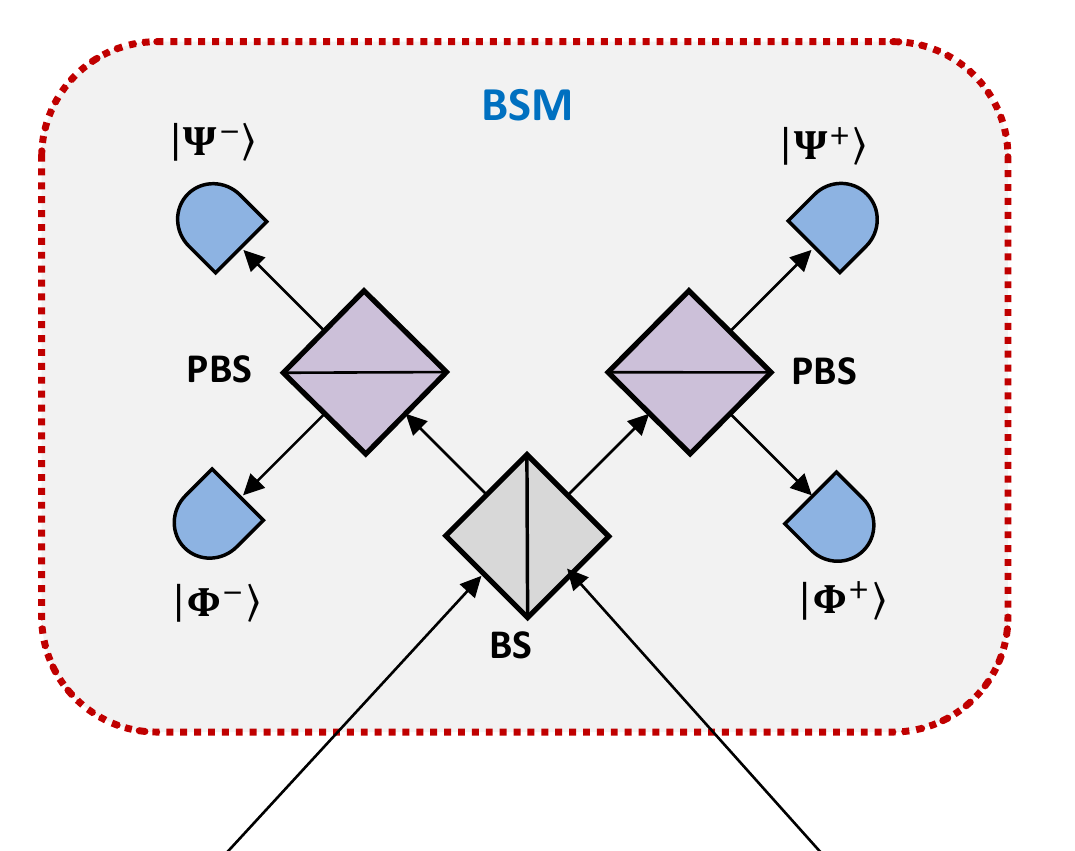}} \caption{Schematic diagram of the Bell state measurement (BSM) device employed in detector-device-independent (DDI) QKD}
\end{figure}

The basic idea of the detector blinding attack is as follows \cite{Lydersen102}. First, Eve sends bright light into Bob's system to force the SPDs into the so-called linear operation mode. In this mode, the detector remains sensitive to bright light with a classical optical power threshold $P_{th}$. Second, Eve performs an intercept and resend attack: she intercepts Alice's signal, measures it in a randomly chosen basis, and resends a bright pulse to Bob according to her measurement result. The peak power of the bright pulse is just above $P_{th}$. If Bob and Eve happen to use the same basis, the bright pulse will be conveyed to one detector and a ``click'' will be registered. Otherwise, the optical power of the bright pulse will be distributed between two SPDs and no detector clicks. Equivalently, Eve has control of Bob's measurement basis and it is easy to show that in principle Eve can learn the whole key without introducing errors.

The detector blinding attack could also be applied in DDI-QKD if there are certain overlooked imperfections in the BSM. Although the actual implementations of the BSM in \cite{Gonzalez14,Lim14,Cao14} are slightly different, all of them employ four SPDs to deterministically identify the four Bell states, $\{|\Psi^{+}\rangle,|\Psi^{-}\rangle, |\Phi^{+}\rangle,|\Phi^{-}\rangle\}$, as shown in Fig.2. In normal QKD operation, if Alice and Bob use the same basis, only two out of the four SPDs have non-zero probability ($50\%$ each in the ideal case) to detect a photon. On the other hand, if Alice and Bob use different bases, all the four SPDs have non-zero probability ($25\%$ each in the ideal case) to click.

Suppose Eve launches the detector blinding attack described above. The optical power of Eve's bright pulse will be distributed either between two SPDs (if Eve and Bob use the same basis) or among four SPDs (if they use different bases). By carefully controlling the power of the bright pulse, Eve can make sure that Bob can register a detection event only when they use the same basis.

The above attack can be easily detected if the beam splitters in the BSM are symmetric and all the SPDs are identical. This is because when Bob and Eve use the same basis, the optical power of Eve's bright pulse will be evenly distributed between two SPDs and result in an unusually high double-click rate \cite{Silva14}. However, if we allow certain uncharacterized imperfections in the BSM, Eve could refine her attack to make it undetectable.  For example, if the SPDs have different wavelength-dependent efficiencies, then Eve could reduce the double-click rate by tailoring the wavelength of her bright pulse. To rule out the possibility of the detector blinding attack in DDI-QKD, we need to quantify the imperfections inside the BSM carefully, or introduce other countermeasures. 

In summary, to improve the security of QKD in practice, various QKD protocols, including DI-QKD, MDI-QKD and DDI-QKD, have been proposed. Each of them has its own advantages and drawbacks, and might be best for certain applications. In this paper, we investigate some underlying assumptions in DDI-QKD. Our results show that if the BSM in DDI-QKD is completely untrusted, a simple attack could allow Eve to gain information of the quantum key without being detected. To prevent the above attack, either countermeasures to Trojan horse attacks or some trustworthiness to the ``untrusted'' BSM device is required. All these details should be clearly specified and included in the security analysis.

I would like to thank Marcos Curty for very helpful discussions.  This work was performed at Oak Ridge National Laboratory, operated by UT-Battelle for the US
Department of Energy under Contract No. DE-AC05-00OR22725.

\end{document}